\documentclass[twocolumn]{aa}
\usepackage{graphicx}
\usepackage{txfonts}
\begin{document}

\newcommand {\flux} {{$\times$ 10$^{-11}$ erg cm$^{-2}$ s$^{-1}$}}
\def\magcir{\raise -2.truept\hbox{\rlap{\hbox{$\sim$}}\raise5.truept
\hbox{$>$}\ }}

 \title{{Using the ROSAT Bright Source Catalogue to find Counterparts for IBIS/ISGRI Survey Sources}}
  \subtitle{}

   \author{ J. B. Stephen\inst{1}, L. Bassani\inst{1}, M. Molina\inst{2}, A. Malizia\inst{1}, A. Bazzano\inst{3}, P. Ubertini\inst{3}, A.J. Dean\inst{4}, A.J. Bird\inst{4}, R. Much\inst{5} and R. Walter\inst{6}}

   \offprints{stephen@bo.iasf.cnr.it}

   \institute{IASF/CNR, Via Piero Gobetti 101, I-40129 Bologna, Italy, 
   \and  Dipartimento di Astronomia, Universit\`a di Bologna, via Ranzani 1, I-40127 Bologna, Italy,
   \and  IASF/CNR, Via del Fosso del Cavaliere, I-00133 Roma, Italy,
   \and  University of Southampton, Southampton, UK,
   \and  ESA-ESTEC, Research and Scientific Support Department, Keplerlaan 1, 2201 AZ, Noordwijk, Netherlands
   \and  INTEGRAL Science Data Centre, Chemin d'Ecogia 16, 1291 Versoix, Switzerland}
   \date{Received / accepted}

  \titlerunning{}

   \authorrunning{J.B. Stephen et al.}

\abstract{
The IBIS/ISGRI first year galactic plane survey has produced a catalogue containing 123 hard X-ray sources
visible down to a flux limit of a few milliCrabs. The point source location accuracy of typically 
1-3 arcminutes has allowed the counterparts for 95 of these sources
to be found at other wavelengths. In order to identify the remaining 28
objects, we have cross-correlated the ISGRI catalogue with the ROSAT All Sky Survey Bright Source Catalogue.
In this way, for ISGRI sources which have a counterpart in soft X-rays, we can use the, much smaller, ROSAT 
error box to search for identifications. As expected, we find a strong correlation between the two
catalogues and calculate that there are 75 associations with the number expected by chance to be almost zero.
Of these 75 sources, ten are in the list of unidentified objects. Using the ROSAT error boxes we provide tentative
associations for 8 of these.

}
 \maketitle

%________________________________________________________________

\section{Introduction}
A key strategic objective of the INTEGRAL mission is a regular survey of
the galactic plane complemented by a deep exposure of the Galactic Centre (Winkler et al. 2003). 
This makes use of the unique imaging capability of IBIS (Ubertini et al. 2003) which
allows the detection of sources at the mCrab level with an angular
resolution of 12 arcminutes and a point source location accuracy (PSLA) of
typically 1-3 arcmin within a large ($29^{\circ}$ x $29^{\circ}$) field of view. 
In its first year, there have been several surveys produced from the 
IBIS/ISGRI data (Revnivtsev et al. 2004a; Molkov et al. 2004; Bird et al. 2004). 
The most complete is that of Bird et al which has detected 123 sources down to a 
flux level of a few mCrab between 20-100 keV. Within this sample of hard X-ray emitting objects,
there are 53 low mass and 23 high mass X-ray binaries, 5 AGN
and various other objects such as pulsars, cataclysmic variables
and a dwarf nova. The remaining objects ($\sim$30 or about 24$\%$ of the
sample) have no obvious counterparts at other wavelengths and therefore
cannot yet be associated with any known class of high energy emitting
objects. Searching for counterparts of these new high energy sources
is of course a primary objective of the survey work but it is made 
very difficult by the good, but still too large, INTEGRAL error
boxes. Cross correlations with catalogues in other wavebands can be used as a 
useful tool with which to restrict the positional uncertainty of the objects detected by IBIS and so to facilitate 
the identification process. Herein, we report on the very strong level
of correlation between the ISGRI first year catalogue and the ROSAT All 
Sky Survey Bright Source Catalogue (RASSBSC). As a result of this work, 
we estimate the current uncertainty in the ISGRI positions and obtain a 
number of possible identifications with objects at other frequencies. Obviously 
this result provides a useful tool with which to validate our source detection 
procedure in future IBIS survey work. Furthermore, the information provided by ROSAT, 
in combination with the IBIS data can be used in population studies as well as for 
diagnostic analysis.\\ 

\section{Cross correlating the IBIS/ISGRI survey and ROSAT Galactic Plane Survey}
The ROSAT all-sky survey (RASS; Voges 1992), performed in the period July 1990 to February 1991,
was carried out with the X-ray telescope
(XRT) and the Position Sensitive Proportional Counter (PSPC,
Pfeffermann et al. 1986). The survey mapped about 60000 sources
in the soft X-ray band (0.1-2.4 keV) down to a limiting flux of the order 
of a few 10$^{-13}$ erg cm$^{-2}$ s$^{-1}$. From this source list 
the Bright Source Catalogue (RASSBSC-1.4.2RXS), containing 18806 RASS sources having a PSPC
count rate larger than 0.05 cts/s in the 0.1-2.4 keV band  and at least 15 source counts,
was extracted providing the positions of the brightest objects at soft X-ray energies 
(Voges et al. 1999). Many of these sources have been identified as stars, AGN, cataclysmic
variables and accreting binaries, most of which are
expected to emit above 10-20 keV, i.e in the IBIS/ISGRI
regime. In order to investigate this, we have cross-correlated the ISGRI first year survey (Bird et al. 2004) with the RASSBSC.

We took both catalogues and calculated the number of ISGRI sources for which at least one
ROSAT counterpart was within a specified distance, out to a maximum of around 3 degrees. 
To have a control group we then created a list of 'anti-ISGRI sources', mirrored in Galactic longitude and latitude, and the same correlation algorithm applied. Figure 1 shows the results of this process.
The lower solid curve is the 'anti-source' correlation, while the dashed line is that expected from
chance correlations given the number of ISGRI objects, the number of ROSAT sources and assuming that the 
latter are evenly distributed across the sky. It is clear that, in this case, the number of correlations can be 
completely explained by chance. The upper solid curve, however, shows the number of associations for the ISGRI 
catalogue, and demonstrates that a strong correlation exists, out to a few hundreds of arcseconds before chance 
coincidences begin to dominate at about 10 arcminutes. In particular, we see that at an error radius of about 
3 arcminutes, the statistical number of chance coincidences expected is 0 (0.35).

\begin{figure}
\centering
\includegraphics[width=8.0cm,height=7.0cm]{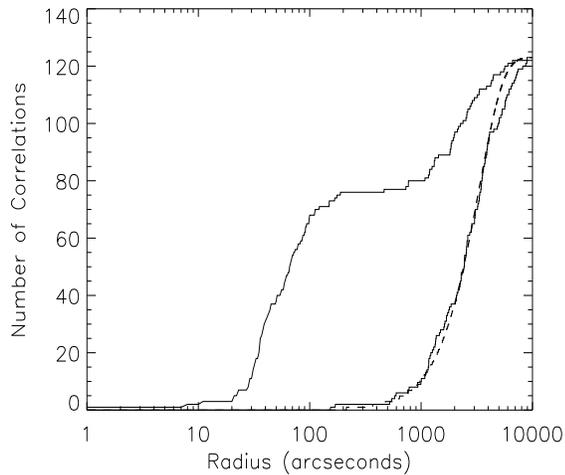}
\caption{The Number of ISGRI/ROSAT (upper solid) and 'Anti-ISGRI'/ROSAT (lower solid) associations as a function of distance. 
The dashed line shows the number of correlations expected by chance.}
\end{figure}

The shape of the figure, shown in more detail in figure 2, reveals two aspects about the ISGRI sources - the curve
should be consistent with the point spread location accuracy (PSLA), as the ISGRI uncertainty in position dominates the
ROSAT error, while the total number of correlations is given by where it flattens off. By fitting a Gaussian function 
to the curve we obtain an average PSLA over all sources of about 1.25 arcminutes (90$\%$ confidence) and a total number 
of 75 correlations. In reality, the PSLA of ISGRI sources is dependent on the source significance, and so the shape 
of the curve would be the sum of many Gaussians of varying widths. By fitting two Gaussians we obtain a significantly 
better fit to the data with an average PSLA of 1 arcminute for sources stronger than about 20$\sigma$ and 2.5 arcminutes
for weaker detections. These values are consostent with those reported by the ISGRI team 
(Gros et al 2003) and indicate that the mosaicing procedure used for summing together such large amounts of data 
does not significantly alter the image quality. 

\begin{figure}
\centering
\includegraphics[width=8.0cm,height=7.0cm]{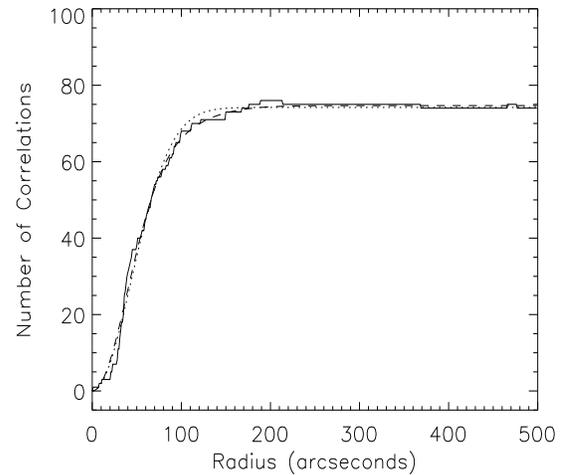}
\caption{A close-up of the ISGRI-ROSAT Correlation distribution. The one-Gaussian (dotted) and two-Gaussian (dashed) fits
to the data are also shown. }
\end{figure}

Despite the strong correlation found between the two catalogues, a significant fraction ($\sim$40\%) of the ISGRI 
sources have no association with a ROSAT object. This may be due to strong absorption preventing detection in soft 
X-rays. Of the overall sample
of 75 ROSAT sources associated with ISGRI detections, 38 are low mass
X-ray binaries, 13 high mass X-ray binaries, 4 AGN/Clusters, 5 pulsars, 3
Cataclysmic variables and 2 other sources, giving a total of 65 identifications.
This leaves a total of 10 ISGRI objects not yet optically identified which statistically should be in the RASSBSC.
Therefore we can now take the ROSAT positional
uncertainty and reduce the ISGRI error box in the search for possible
counterparts; in some cases ROSAT error box is of the order of 6-10
arcseconds which is sufficiently small to highlight one or two likely
counterparts. \\

\section{Searching for counterparts of unidentified IBIS/ISGRI sources} 

For the 10 IBIS survey sources which are unidentified and  have  a possible counterpart in the RASSBSC, Table 1
reports the ISGRI name, ROSAT coordinates and error box radius as well as the
distance of the ROSAT position from that of the IBIS/ISGRI best fit
source location. The ROSAT uncertainty provides a smaller 
error box than ISGRI and so allows an easier  search for counterparts. In the following, we provide details 
on this search and indicate a number of optical counterparts. In some cases, 
follow up spectroscopic observations  of these counterparts have already permitted secure identifications.\\

\begin{table*}
\begin{center}
\centerline{{\bf Table 3: Unidentified ISGRI Sources with a RASSBSC counterpart}}
\begin{tabular}{rccc}
\hline\hline
{\bf Name} & {\bf ROSAT Coord.} & {\bf Positional error($''$)} & {\bf Distance ($'$)} \\ 
\hline
IGR J$15479-4529$ & 15 48 14.50 $-45$ 28 45.0 &  9  & 0.54  \\
IGR J$16167-4957$ & 16 16 37.20 $-49$ 58 47.5 & 16  & 1.35  \\
IGR J$16558-5203$ & 16 56 05.60 $-52$ 03 45.5 &  8  & 0.97  \\
4U$1705-32$       & 17 08 54.40 $-32$ 18 57.5 &  8  & 1.19  \\
IGR J$17303-0601$ & 17 30 21.50 $-05$ 59 33.5 &  7  & 0.44  \\
IGR J$17195-4100$ & 17 19 35.60 $-41$ 00 54.5 &  8  & 0.89  \\
IGR J$17200-3116$ & 17 20 06.10 $-31$ 17 02.0 &  9  & 1.15  \\
IGR J$17254-3257$ & 17 25 25.50 $-32$ 57 17.5 & 14  & 1.23  \\
IGR J$17488-3253$ & 17 48 54.71 $-32$ 54 44.0 & 11  & 1.61  \\
XTE J$1901+014$     & 19 01 41.00 $+01$ 26 18.5 & 12  & 2.80  \\
\hline
\end{tabular}
\end{center}
\end{table*}

{\bf IGR J$\bf15479-4529$}
The identification of IGR J$15479-4529$  can be taken as a case study: 
this source was discovered during an observation of the X-ray transient 4U $1630-47$ (Tomsick et al. 2004)
and localized at RA (J2000) = 15 47 0.9 and Dec (J2000) = $-45$ 29 00 ($2'$ uncertainty); 
this position is $3.6'$ from the location of 1RXS J$154814.5-452845$, a bright ROSAT all sky survey source recently
identified with an intermediate polar cataclysmic variable (Haberl et al. 2002).  
Although the ROSAT  source was
outside the ISGRI error box, observations by XMM-Newton indicate that
the object is a strong X-ray emitter up to 10 keV , suggesting the possibility of an association between the 
X and gamma-ray emissions.  The IBIS/ISGRI first year survey has
been able to provide a more refined position, locating the gamma-ray source only $0.7'$ away from 
1RXS J$154814.5-452845$ thus confirming the association  and therefore providing a secure
identification for this IBIS object.

{\bf IGR J$\bf16167-4957$}
This is one of the only two survey sources located in the Norma region and associated with 
a bright ROSAT object (the other being IGR J$16558-5203$); this is possibly due to a combination 
of higher flux and lower extinction which prevents soft X-rays being totally absorbed in these two cases. 
Within the ROSAT error box of  IGR J$16167-4957$ there are too many optical and/or near-infrared counterparts 
in the USNO-B1/2-MASS catalogues (Monet et al. 2003; Cutri et al. 2003) for a fruitful 
identification search; clearly an error box larger than 10 arcsec is of no use in crowded regions of the 
galactic plane/centre.
 
{\bf IGR J$\bf16558-5203$}
This is the other  survey source located in the Norma Region and associated with a 
bright ROSAT source; by itself the  X-ray error box, although small, is insufficient to
pinpoint a single candidate. Luckily the soft X-ray source has also been observed  with
the ROSAT/HRI instrument and detected as RXH J$165605.6-520339$; in this
case the uncertainty associated with the position is of the order of a few arcseconds, which is
sufficient to pinpoint a likely optical/infrared candidate: a USNO-B1 source located at 
RA=16 56 05.6 and Dec=$-52$ 03 40.4. This, possibly extended, source is quite bright in both B and R optical 
bands (B$\sim$14.5, R$\sim$13.5) and is also visible in the near infrared (2-MASS). 
Follow up optical spectroscopic observations of this likely counterpart can prove its association with the 
IBIS source and define its nature. 

{\bf 4U$\bf1705-32$}
Although previously detected both by the Uhuru and HEAO satellites in
the 2-6 keV band (2-45 10$^{-11}$ erg cm$^{-2}$ s$^{-1}$, Forman et al. 1978 and Wood et al. 1984) 
and despite its X-ray brightness, this source is
still unidentified. The ROSAT PCPC error box provides two optical 
counterparts, of which only one is also a near infrared source 
(2-MASS and DENIS).  Here too the source error box has been
observed with the HRI instrument on various occasions: the source was found to be
variable by at least a factor of 6 over a six month period. Inspection of the small HRI error 
circle  (at RA=17 08 54.23 Dec=$-32$ 18 55.8 with a $2''$ radius) indicates the lack of any optical and/or 
infrared conterpart suggesting that the source is extremely weak (or highly absorbed) in optical with an 
upper limit on B of $\sim$20 magnitudes. Within $5''$ from the ROSAT/HRI position lies an USNO B-1 source at 
RA=17 08 53.85 Dec=$-32$ 18 55.4 having B$\sim$18 and I$\sim$16; no source is listed at this position in the 
2-MASS catalogue. Follow-up work on this nearby object and a deep search of the HRI error could shed light on this
bright X-ray emitter.

{\bf IGR J$\bf17303-0601$}
This source coincides with the HEAO A1 object 1H$1726-058$ having a 2-10 keV flux of 
1.5 10$^{-11}$ erg cm$^{-2}$ s$^{-1}$ (Wood et al. 1984).  
An object close to the ISGRI position  is also listed in the recent
RXTE all sky survey (XSS$17309-0552$, see Revnivtsev et al. 2004b). 
Sazanov and Revnivtsev (2004) suggested an extragalactic origin  on
the basis of its similarity with the spectral slope of this class of sources as detected by RXTE.
Inspection of the ROSAT error box indicates the presence of
two likely counterparts both with detection at near-infrared (2-MASS survey) and optical frequencies 
(USNO-B1 catalogue).  The furthest to the ROSAT position is also the brightest of the two
sources (R$\sim$15.5 ). 
Recent optical spectroscopy of both candidates has allowed us to identify the true counterpart and also to 
identify it with a low mass-X-ray binary system (Masetti et al. 2004 and details therein). 

{\bf IGR J$\bf17195-4100$}
The ROSAT error box of this ISGRI source contains only one bright (B$\sim$15) optical source
according to the USNO-B1 catalogue (RA=17 19 35.94 Dec=$-41$ 00 53.5). Within $2.2''$ from this source lies a 
bright near-infrared source which is possibly extended or unresolved in the direction of the USNO-B1 object.
Two more optical sources are located just outside the ROSAT error circle. 
Clearly follow up optical  work on  the most likely counterpart (and possibly of the nearby objects too) can 
determine if the source has the characteristics of a high energy emitting source and  further determine its 
association to the near-infrared object.  

{\bf IGR J$\bf17200-3116$}
Within the ROSAT error box there is only one USNO B1 source at RA=17 20 06.1 Dec=$-31$ 17 02.0
($\sim$19 magnitude in B) : this source is also near-infrared detected. However, inspection of the
2-MASS survey indicates the presence of two extra near-infrared objects. Although these two objects are 
brighter than the USNO-B1 source in JHK band photometry,  they must be  extremly reddened in optical as 
they are below the USNO-B1 catalogue threshold. 

{\bf IGR J$\bf17254-3257$}
Inspection of the ROSAT error box indicates the presence of only one optical source in USNO-B1 at
RA=17 25 26.03 Dec=$-32$ 57 06.1; this source has a B magnitude in the range 18-19 and is also 
near-infrared detected. However, the 2-MASS catalogue reports 5 more sources within the X-ray 
error box, again a clear indication that an  error radius greater than 10 arcsec 
makes the identification procedure highly difficult.

{\bf IGR J$\bf17488-3253$}
The positional uncertainty on this ROSAT source is too large for a
fruitful identification process; luckily this source was also
observed by the HRI instrument (1RXH J$174854.9-325448$) and localized
at RA=17 48 54.9 Dec=$-32$ 54 48.2 with an uncertainty of a few arcseconds. Within this error box
no optical, radio and/or infrared counterpart is found, implying a very weak source at
all these frequencies, however at the edge of the ROSAT error circle a (possibly extended) source with 
R$\sim$16 is clearly seen on DSS2. Optical photometry can disintangle whether this object is 
extended or made of a group of point like sources and spectroscopy will be able to pinpoint the nature 
of the source(s) responsible for the high energy emission.

{\bf XTE J$\bf1901+014$}
This source has often been detected at soft X-ray energies 
by Einstein (2E1859.1+0122, Hertz \& Grindlay 1988), EXOSAT (GPS1858+015, Warwick et al. 1988)
and most recently by ROSAT.  Apart from the survey detection, the source was also seen during a pointed ROSAT/HRI 
measurement performed on 1994 October 3, however, its position appears  to be slightly offset from the 
RASSBSC coordinates 
(the HRI position is RA= 19 01 40.1 Dec=+01 26 30.6 with an error radius of $10''$, 
Wijnands 2002). Although these locations are compatible within the errors, 
each contains various possible counterparts: within the RASSBSC error box (see Table 1)
lies a single object in the USNO-B1/2-MASS catalogue at RA=19 01 41.11  and Dec=01 26 28.3 
having B, R and K magnitudes of 17 and 14.5 and 13.7 respectively. Instead within the
HRI error box we only find a 2-MASS object at RA=19 01 39.84 and Dec=+01 26 32.6 
with K=10.4 and no optical counterpart. 

\section{Conclusions}

We have shown that, as expected, there is a strong correlation between the ISGRI survey source 
list and the ROSAT All Sky Survey  Bright Source Catalogue.
By analysing this correlation, we have further shown that the ISGRI positional uncertainty is 
similar for the mosaiced images as for the 
individual observations at around 1-3 arcminutes. Furthermore the correlation function allows 
us to calculate that there should be 75 clear 
ISGRI-ROSAT source correlations (with the number due to chance being almost zero), and as only 65 have already been identified, we can use this 
information to help find optical counterparts 
for the other 10 sources by using the associated ROSAT positional
uncertainty. In eight cases the ROSAT error box is sufficiently small to pinpoint 1-3 likely
counterparts. We can also use the ROSAT
associations to validate source detections in future survey work when,
working at lower significance levels, we expect to introduce spurious
sources due to imperfect image cleaning: a likely ROSAT counterpart
would increase the probability that any excess found is a real
IBIS/ISGRI source. Population studies, as well as diagnostic analysis of the ROSAT/IBIS-ISGRI associations
is in progress and will be presented in a future work.\\

\begin{acknowledgements}
This research has been partially supported by ASI contract I/R/041/02.
This analysis has made use of the HEASARC archive which is a service of the Laboratory for High Energy Astrophysics (LHEA) at NASA/ GSFC
and the High Energy Astrophysics Division of the Smithsonian Astrophysical Observatory (SAO)
\end{acknowledgements}

\end{document}